\documentclass[preprint,showpacs,preprintnumbers]{revtex4}
\usepackage{graphicx}
\usepackage{dcolumn}
\usepackage{bm}
\usepackage{epstopdf}
\usepackage{amsmath}

\newcommand{\be}{\begin{equation}}
\newcommand{\en}{\end{equation}}
\newcommand{\bea}{\begin{eqnarray}}
\newcommand{\ena}{\end{eqnarray}}
\begin{document}

\title{ The  generalized second law of thermodynamics for  interacting $f(R)$ gravity}
\author{Ram\'on Herrera}
\email{ramon.herrera@ucv.cl}
\affiliation{Instituto de F\'{\i}sica, Pontificia Universidad Cat\'{o}lica de Valpara%
\'{\i}so, Casilla 4059, Valpara\'{\i}so, Chile.}
\date{\today }
\author{Nelson Videla}
\email{nelson.videla@ucv.cl}
\affiliation{Instituto de F\'{\i}sica, Pontificia Universidad Cat\'{o}lica de Valpara%
\'{\i}so, Casilla 4059, Valpara\'{\i}so, Chile.}

\begin{abstract}
We examine  the validity of the generalized second law (GSL) of
gravitational thermodynamics in the context of  interacting
$f(R)$ gravity. We take into account that the boundary of the
universe to be confined  by the dynamical apparent horizon in a
flat FRW universe. We study the effective equation of state,
deceleration parameter and GSL in this interaction-framework. We
find that the evolution of the total entropy increases through the
interaction term.   As a example, we consider a $f(R)$ gravity
with a power-law dependence on the curvature $R$. Here,  we find
exact solutions for a model in which the interaction term is
related to the total energy density of matter.
\end{abstract}

\maketitle



\section{Introduction}

The observational data of the luminosity-redshift of type Ia
supernovae (SNeIa), large scale structure (LSS) and the cosmic
microwave background (CMB) anisotropy spectrum, have supported
evidence that our universe has recently arrived a phase of
accelerated expansion \citep{plck2013, wmap2012, perm2003,
perm1999, gar1998, rie1998}. For this current acceleration  a
possible responsible is the dark energy (DE)  and the nature of
this DE is a problem today. For a review of DE  models, see Refs.
\citep{li2011, peb2003, pad2003, st1999, zla1999, cal1998}.

In the last years, a $f(R)$ theory was proposed  to elucidate  the
expansion of the universe without taking  the DE \citep{def2010,
hu2007, noj2006}. In such an approach, the Ricci scalar $R$ in the
Einstein-Hilbert action is replaced by a general function $f(R)$,
(for a review see Refs. \citep{spo2014, noj2011,sot2010, am2007}).
Also, there are other classes of modified gravities that can give
an explication for the different cosmological scenarios without
take into account to the DE. In particular, $f(T)$ theory that is
a generalization of the teleparallel gravity (TG) and becomes
equivalent of General Relativity \citep{haya1979,
eins1928,Bamba:2014eea}. Also, the modified gravity that includes
the Gauss-Bonnet invariant term $f(\mathcal{G})$
\citep{Nojiri:2005jg,Cognola:2006eg,mod2012, felice2009, tsu2002,
car2000, ant1994}.  In this context, there are two forms to
analyze dark energy energy models, by means of a fluid explanation
and the other is to define the action associate to a scalar field
(see e.g. Ref.\cite{Bamba:2012cp}). In particular, for the
background solutions these two ways to describe dark energy models
are equivalent. However, one cannot assume univocally this
equivalence, for example  in the stability of the
solutions\cite{C1} or in the studies of cosmological
perturbations\cite{def2010,C2}.


On the other hand, in the context of the thermodynamics point of
view, the accelerating universe has conceived  much
consideration  and different types of  consequences  has been
detected \citep{haw1975, bek1972}. Specifically, the confirmation  of the first
and second law of the thermodynamics, from  a  dynamic aspect
together with a thermodynamic analysis of the accelerating
universe.

For the validity of the  generalized second law (GSL) of
thermodynamics, is essential  that the evolution with respect to
the cosmic time of the total entropy $\dot{S}_{Total}$, becomes
$\dot{S}_{Total}=d(S_A+S_m)/dt\geq 0$.  $S_A$ is the
Bekenstein-Hawking entropy on the apparent horizon and $S_m$
represents the entropy of the universe filled with matter \citep{cai2005}.
Consequently, in accordance with the GSL of thermodynamics, the
development  of the total entropy, $S_{Total}$, cannot decrease in
the time \citep{bru2000, pav1990, pau1988, pau1987, bek1974, bek1973}.

In the frame of reference of the first law of thermodynamics, we
can write for the apparent horizon $-dE=T_A\,dS_A$  and obtain in
this form the Einstein's field equation. Also, the agrement
between the first law of thermodynamics and the Einstein's field
equation is fulfilled, if we take into account that the Hawking
temperature $T_A$, in which $T_A\propto R_A^{-1}$ and also the
entropy on the apparent horizon  $S_A \propto A$, in which  $R_A$
and $A$ are the radius and area related to the horizon
\citep{cai2005}, see also Ref. \citep{jam2010, she2009, akb2007}.
Nevertheless, it should also be note that the entropy on the
apparent horizon, is modified for other types of theories.
Specifically, in  $f(R)$ gravity, the geometric entropy is given
by $S_A=Af_R/4G$ \citep{akb2006}, where $f_R=\partial
f(R)/\partial R$. Also, in the context of the model $f(T)$
gravity, in Ref. \citep{mia2011}, the authors calculated that when
$f''$ is small, the entropy of the apparent horizon is given by
$S_A=A\,f'/4G$, here the primes denote derivative with respect to
the torsion scalar $T$, see also Ref.\cite{p1}.

In the framework of $f(R)$ gravity  the GSL of thermodynamics, was
studied  in Ref. \citep{kar2012} (see also Ref. \citep{bam2009, akbar2007, eli2006}). In this
work, the authors analyzed a Friedmann-Robertson-Walker (FRW)
universe filled only with ordinary matter enclosed by $S_A$ and
examined  the validity of the GSL for a viable $f(R)$ model;
$f(R)=R-\alpha/R+\beta R^2.$

On the other hand,  to solve the cosmic coincidence problem
\citep{peb2003,st1997}, several authors  have analyzed  the
interaction between DE and DM components \citep{zim2001, bil2000,
amen2000, ame2000}. Here, the interaction term can mitigate the
coincidence problem in the sense that the rate between both
densities  either leads to a constant or changes slowly in late
times \citep{sdc2009, s2008}. In connection with the GSL, the
analysis  of the validity of the GSL in the presence of an
interaction between DM and DE was studied in Ref. \citep{k2010}.
In this model, the authors considered that the interaction between
both component is proportional to the DE. Additionally, the
thermodynamic description for the interaction between holographic
DE and DM was considered in Ref. \citep{wang2008} and also an
analysis of the GSL for the interacting generalized Chaplygin gas
model was studied  in \citep{kmi2011}. In the context of the
interaction between DE and DM from the Le Ch\^{a}telier-Braun
principle  was analyzed in Ref. \citep{dpav2009}.

The goal of this work is to study the validity of the GSL of thermodynamics
considering  the interacting $f(R)$ gravity model. We will analyze a flat universe  FRW
background filled with the pressureless matter. Also, we study
the equation of state (EoS) of the model, the deceleration
parameter, and the GSL of gravitational thermodynamics. Finally as
an example, we analyzed  a $f(R)$ model together with a particular
interaction term.

The outline of the paper is as follows. The next section presents
the interacting $f(R)$ gravity in a flat FRW universe. Here, we
investigate the EoS and the deceleration parameter. Section
\ref{g} we study the validity of the GSL of thermodynamics in the
context of the interacting $f(R)$ gravity. Section \ref{e} we
analyze an example for $f(R)$ and a particular  interaction term
$Q$. Section \ref{co} we study the conformal transformation and
the GSL in a scalar tensor gravity theory.
 Finally, in Sect.\ref{conclu} we summarize our finding. We
chose units such that $c=\hbar =8\pi G=1$.

\section{ Interacting $f(R)$ gravity}\label{f}

The action $I$  in the framework of $f(R)$ gravity, becomes
\citep{noj2011, soti2010, bmb2009}
\begin{equation}
I=\int\,d^4x\,\sqrt{-g}\,\left[\frac{f(R)}{2}+L_m\right].\label{action}
\end{equation}
Here $L_m$ is related to the Lagrangian density of the matter
inside the universe.

In order to describe the $f(R)$ theory we start with  the
following gravitational field equations in a flat FRW background
filled with the pressureless matter

\begin{equation}
H^2=\frac{1}{3}\,\rho_t,\label{H}
\end{equation}
\begin{equation}
\dot{H}=-\frac{1}{2}\,(\rho_t+p_t),\label{dH}
\end{equation}
where $\rho_t$ and $p_t$ are the total energy density and pressure
given by
\begin{equation}
\rho_t=\frac{\rho_m}{f_R}+\rho_R\,,\,\,\,\,\,\mbox{and}\,\,\,\,\,\,\,p_t=p_R.\label{rhot}
\end{equation}
Here, $\rho_R$ and $p_R$ are the energy density and pressure due
to the curvature contribution, defined as \citep{noj2011,
soti2010, bmb2009}
\begin{equation}
\rho_R=\frac{1}{f_R}\,\left(-\frac{1}{2}(f-Rf_R)-3H\dot{f}_R\right),\label{rhoT}
\end{equation}
and
\begin{equation}
p_R=\frac{1}{f_R}\,\,\left(\frac{1}{2}(f-Rf_R)+2H\dot{f}_R+\ddot{f}_R\right),
\end{equation}
and the energy density of the matter $\rho_m$, is given by
\begin{equation}
\rho_m=\frac{f}{2}-3\left(\dot{H}+H^2-H\frac{d}{dt}\right)f_R\,,\label{rhom}
\end{equation}
where,  $H=\dot{a}/a$ is the Hubble factor, $a$ is a scale factor,
$R=6\left(\dot{H}+2H^2\right)$ is the scalar curvature  and
$f_R=\partial f(R)/\partial R$. Dots here mean derivatives with
respect to the cosmological time.

On the other hand, we shall consider  that both components, i.e.,
the scalar curvature  and the cold dark matter do not conserve
separately but that they interact through a $Q$ term (to be
specified later) according to

\begin{equation}
\dot{\rho_m}+3H\rho_m=\,Q,\label{drm}
\end{equation}
and
\begin{equation}
\dot{\rho_R}+3H(\rho_R+p_R)-\frac{\dot{f}_R}{f_R^2}\rho_m=\,-Q.\label{rt}
\end{equation}
Note that the energy conservation law for the total perfect fluid
is $\dot{\rho}_t+3H(\rho_t+p_t)=0$. In what follows we shall
assume $Q>0$. We also consider that the curvature contribution
component obeys an equation of state (EoS) parameter
$w_R=p_R/\rho_R$ and then the Eq.(\ref{rt}), becomes

\begin{equation}
\dot{\rho_R}-\frac{\dot{f}_R}{f_R^2}\rho_m+3H\rho_R\left(1+w_R+\frac{Q}{3H\rho_R}\right)=0.\label{rt2}
\end{equation}

Taking time derivative of Eq.(\ref{rhoT}), we obtain
\begin{align}
&   \dot{\rho_R}=-\frac{\dot{f}_R}{f^2_R}\,\left(-\frac{1}{2}(f-Rf_R)-3H\dot{f}_R\right)\nonumber\\
&   +\frac{1}{f_R}\,\left(\frac{R\dot{f}_R}{2}-3\dot{H}\dot{f}_R-3H\ddot{f}_R\right).\label{ddot}
\end{align}

From Ref.\citep{kmi2011}, we combining Eq.(\ref{rt2}) and (\ref{ddot})
and the EoS parameter results
\begin{equation}
w_R=-\left[1+\frac{Q}{3H\rho_R}+\left(\frac{(\ddot{f}_R-H\dot{f}_R)}{[(f-Rf_R)/2+3H\dot{f}_R]}\right)\right],\label{wt2}
\end{equation}
here, we noted that the Eq.(\ref{wt2}) corresponds to  an
effective EoS parameter.



On the other hand, the deceleration parameter $q$ is defined  as $
q=-\left[1+\frac{\dot{H}}{H^2}\right]$,  and considering
Eqs.(\ref{H}) and (\ref{dH}), yields

\begin{equation}
q=\frac{1}{2}\,\left[1+\frac{\rho_R\,w_R}{H^2}\right].\label{qq}
\end{equation}
Combining Eqs.(\ref{wt2}) and (\ref{qq}) the deceleration
parameter $q$ can be written as
\begin{align}
&   q=\frac{1}{2}+\frac{1}{2H^2\,f_R}\,\,\left(\frac{1}{2}(f-Rf_R)+3H\dot{f}_R\right)\nonumber\\
&   -\frac{1}{6H^3}\left(Q-\frac{3H(\ddot{f}_R-H\dot{f}_R)}{f_R}\,\right).
\label{q}
\end{align}
We noted that for the particular case in which non-interacting
limit $Q=0$ and  $f(R)=R$ the Eq.(\ref{q}) results in $q=1/2$,
representing  to the matter dominated epoch.

\section{GSL interacting - $f(R)$}\label{g}

It is well known  that for the GSL, the entropy of the horizon
plus the entropy of the matter within the horizon cannot decrease
in time, see Refs.\citep{bru2000, pav1990, pau1988, pau1987, bek1974, bek1973}. We consider that the boundary of
the universe to be enclosed by the dynamical apparent horizon in a
flat FRW universe. In this form, the radius of the apparent
horizon $R_A$ coincides with the Hubble horizon and is given by
\citep{hay1996, poi1990}
\begin{equation}
R_A=\frac{1}{H}\,.\label{RA}
\end{equation}

On the other hand,  the Hawking temperature on the apparent
horizon $T_A$ as function of the  radius $R_A$ is defined as
\citep{cai2005}
\begin{equation}
T_A=\frac{1}{2\pi\,R_A}\,\left(1-\frac{\dot{R_A}}{2HR_A}\right),\label{TAA}
\end{equation}
where the ratio $\dot{R_A}/2HR_A<1$, guarantees  that the Hawking
temperature $T_A>0$.

From the Gibb's equation, the entropy of the universe assuming
that the DM inside the apparent horizon, is given by \citep{izq2006}
\begin{equation}
T_A\,dS_m=dE_m+p_m\,dV=dE_m.\label{TA}
\end{equation}
Here, $E_m=V\,\rho_m$ where the volume of the pressureless matter
is defined as $V=4\pi\,R_A^3/3$, then
\begin{equation}
E_m=V\,\rho_m=\frac{4\pi\,R_A^3}{3}\,\rho_m.\label{Em}
\end{equation}
Combining Eqs.(\ref{drm}), (\ref{TA}) and (\ref{Em}), we find
\begin{equation}
T_A\,\dot{S}_m=4\pi\,R_A^2\,\rho_m\left(\dot{R}_A+H\,R_A\left[\frac{Q}{3H\rho_m}-1
\right]\right),\label{Sm}
\end{equation}
where $\dot{S}_m$ correspond to the time derivative of the entropy
from the matter source inside the horizon. Note that in the
non-interacting limit i.e., $Q=0$ the Eq.(\ref{Sm}) reduces to the
standard  Gibb's equation
$T_A\,\dot{S}_m=4\pi\,R_A^2\,\rho_m\left(\dot{R}_A-H\,R_A\right)$.
Also, we observe that the evolution of the matter entropy
$T_A\dot{S}_m$ increases with the introduction of the interaction
term $Q$.

Using  Eqs.(\ref{rhom}) and (\ref{Sm}), we get
\begin{align}
&   T_A\,\dot{S}_m=2\pi\,R_A^2\,[f-6(\dot{H}+H^2-Hd/dt)f_R]\times\nonumber\\
&   \left(\dot{R}_A+
H\,R_A\left[\frac{2\,Q}{3H[f-6(\dot{H}+H^2-Hd/dt)f_R]}-1\right]\right).\label{Sm2}
\end{align}
Here,  as before we note that in the limit $Q=0$, Eq.(\ref{Sm2})
reduces to expression obtained in Ref.\citep{kar2012}, in which
$T_A\dot{S}_m=2\pi\,R_A^2[f-6(\dot{H}+H^2-Hd/dt)f_R]\,(\dot{R}_A-HR_A)$.

On the other hand, the addition  of the apparent horizon entropy
$S_A$, in the framework of $f(R)$ gravity, is given by \citep{odin2005, wald1993}
\begin{equation}
S_A=\frac{A\,f_R}{4\,G},\label{SAA}
\end{equation}
where  the area of the horizon $A$ is defined as $A=4\pi\,R_A^2$.

Taking time derivative  of the above equation and considering
Eq.(\ref{TAA}), the evolution of horizon entropy, can be written
as
\begin{equation}
T_A\,\dot{S}_A=4\pi\left(1-\frac{\dot{R}_A}{2HR_A}\right)\;
\left(2\,\dot{R}_A\,f_R+R_A\,\dot{f}_R\right).\label{SA}
\end{equation}

We note that Eq.(\ref{SA}) coincides with the evolution of the
horizon entropy $T_A\,\dot{S}_A$ estimated in Ref. \citep{kar2012}.
Also, we observe  that $T_A\,\dot{S}_A$ becomes independent of
the interacting term $Q$.

In this form,  the total entropy $S_{Total}$ due to different
contributions of the apparent horizon entropy and the matter
entropy, i.e., $S_{Total}=S_A+S_m$, from Eqs.(\ref{Sm2}) and
(\ref{SA}), becomes
$$
T_A\dot{S}_{Total}=2\pi R_A^2 \,
\Bigl[\,\left(\frac{2}{R_A^2}-\frac{\dot{R}_A}{HR_A^3}\right)\;
\left(2\,\dot{R}_A\,f_R+R_A\,\dot{f}_R\right)\,\,
$$
\begin{align}
&   +[f-6(\dot{H}+H^2-Hd/dt)f_R]\times\nonumber\\
&   \left(\dot{R}_A+
H\,R_A\left[\frac{2\,Q}{3H[f-6(\dot{H}+H^2-Hd/dt)f_R]}-1\right]\right)\,\Bigr].\label{ST}
\end{align}
Note that the interacting-term $Q$ modifies the evolution of the
total entropy,  in which the GSL of thermodynamic, increases by a
factor $4\pi\,R_A^3\,Q/3>0$. Also, we note that in the special
case in which $f(R)=R$, the GSL from Eq.(\ref{ST}) results in
$T_A\,\dot{S}_{Total}=\pi R_A^3[R_A\,\rho_m^2+4Q/3]>0$. In
particular, in the limit $Q=0$ and $f(R)=R$ we obtained
$T_A\,\dot{S}_{Total}=\pi R_A^4\,\rho_m^2>0$ and coincides with
the GSL obtained in Ref. \citep{kar2012} (recalled, that $8\pi\,G=1$).

In the following, we will analyze analytical solutions for the GSL
of thermodynamics for  one specific interaction term $Q$ and a
particular   $f(R)$ gravity model.

\section{An example for $Q$ and $f(R)$: Analytical solutions}\label{e}

Let us consider that the interaction term $Q$ is related to the
total energy density of matter and takes the form \citep{guo2007, rosen2007}
\begin{equation}
Q=3\,c^2\,H\,\rho_m,\label{QQ}
\end{equation}
where $c^2$ is a  positive definite constant and the factor 3 was
considered for mathematical convenience (for a review of Q-terms
see Ref. \citep{bolo2013}).

Inserting the interaction term Q given by Eq.(\ref{QQ}) in the
energy equation of the matter given by  Eq.(\ref{drm}), we find
\begin{equation}
\rho_m=\rho_{m0}\,a^{-3(1-c^2)}\label{solrm}.
\end{equation}

On the other hand,   we study the power-law $f(R)$ model, as a
specific case, where
\begin{equation}
f(R)=\alpha\,R^n\label{fT2},
\end{equation}
in which  $0<n<1$ and $\alpha>0$ are
constants \citep{noj2011,sot2010, am2007}. Here, $n$ is the slope of the
gravity Lagrangian and $\alpha$ with the dimensions taken in such
a way to give $f(R)$ the correct physical dimensions. The model
$R^n$ gravity, like any $f(R)$ theory, is object to experimental
constraints. In this context, in Ref. \citep{co2006}, the authors analyzed
the gravitational lensing in $R^n$ gravity.  In Ref. \citep{zakh2006} was
studied the solar system constraints for $R^n$ model. Also
recently, the constraints on $R^n$ gravity from precession of
orbits of S2-like stars was considered in Ref. \citep{borka2012} (see also
Refs. \citep{cpz2008, tjkw2008, zakh2006, clif2006, barcli2005}).

In this form combining Eqs.(\ref{rhom}), (\ref{solrm}), and
(\ref{fT2}), we get
\begin{equation}
a(t)\propto t^{\frac{2n}{3(1-c^2)}},\label{sol1}
\end{equation}
where the exponent in the scalar factor is $\frac{2n}{3(1-c^2)}>1$
for guarantee an accelerated phase of the universe. Considering
that $0<n<1$, together with the condition $\frac{2n}{3(1-c^2)}>1$,
we get that the range for the parameter $c^2$, becomes
$\frac{(3-2n)}{3}<c^2<1$.

The Hubble parameter is given by
\begin{equation}
H(t)=\frac{2n}{3(1-c^2)}\,\frac{1}{t}=\sqrt{\frac{n\,R}{3(4n-3[1-c^2])}},
\end{equation}
and the acceleration parameter $q$, from Eqs.(\ref{qq}) and
(\ref{sol1}) is given by $q=(3-2n-3c^2)/2n$.

The total entropy  due to different contributions of the apparent
horizon entropy and the matter entropy from Eq.(\ref{ST}), can be
written as
\begin{align}
&   T_A\dot{S}_{Total}=2\pi R_A^2\,\Bigl[\,\left(\frac{2}{R_A^2}-\frac{\dot{R}_A}{HR_A^3}\right)\times\;\nonumber\\
&   \left(2\,\dot{R}_A\,R^{n-1}+(n-1)R^{n-2}R_A\,\dot{R}\right)\alpha
n\,\,\nonumber\\
&   +\alpha[ R^n-6 n(\dot{H}+H^2-Hd/dt)R^{n-1}]\nonumber\\
&   \times\left(\dot{R}_A+
H\,R_A\left[c^2-1\right]\right)\,\Bigr],\label{ST2}
\end{align}
where $R_A=1/H=R^{-1/2}\,\sqrt{\frac{3(4n-3[1-c^2])}{n}}$.

\begin{figure}[th]
\includegraphics[width=3.3in,angle=0,clip=true]{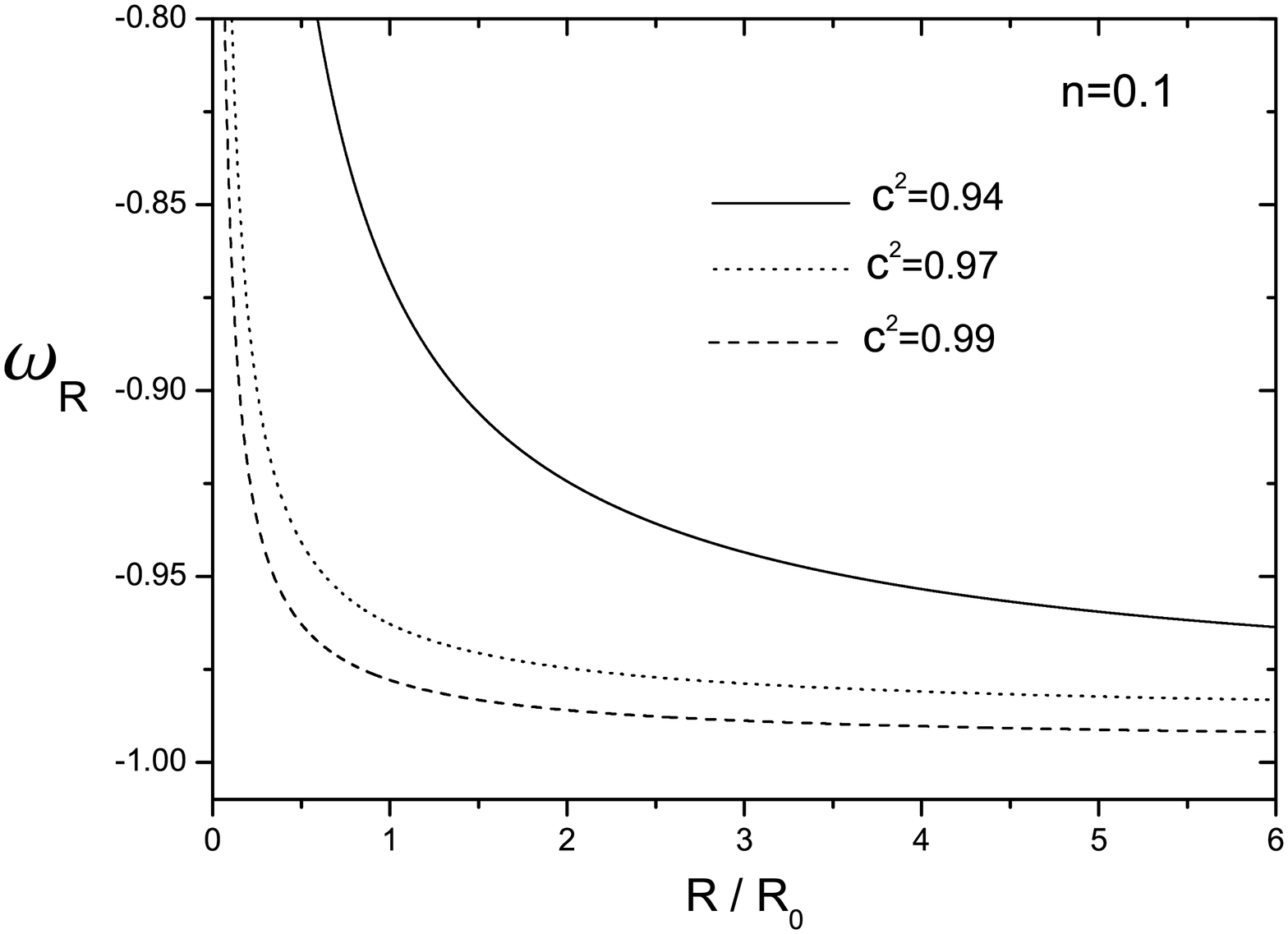}
\includegraphics[width=3.3in,angle=0,clip=true]{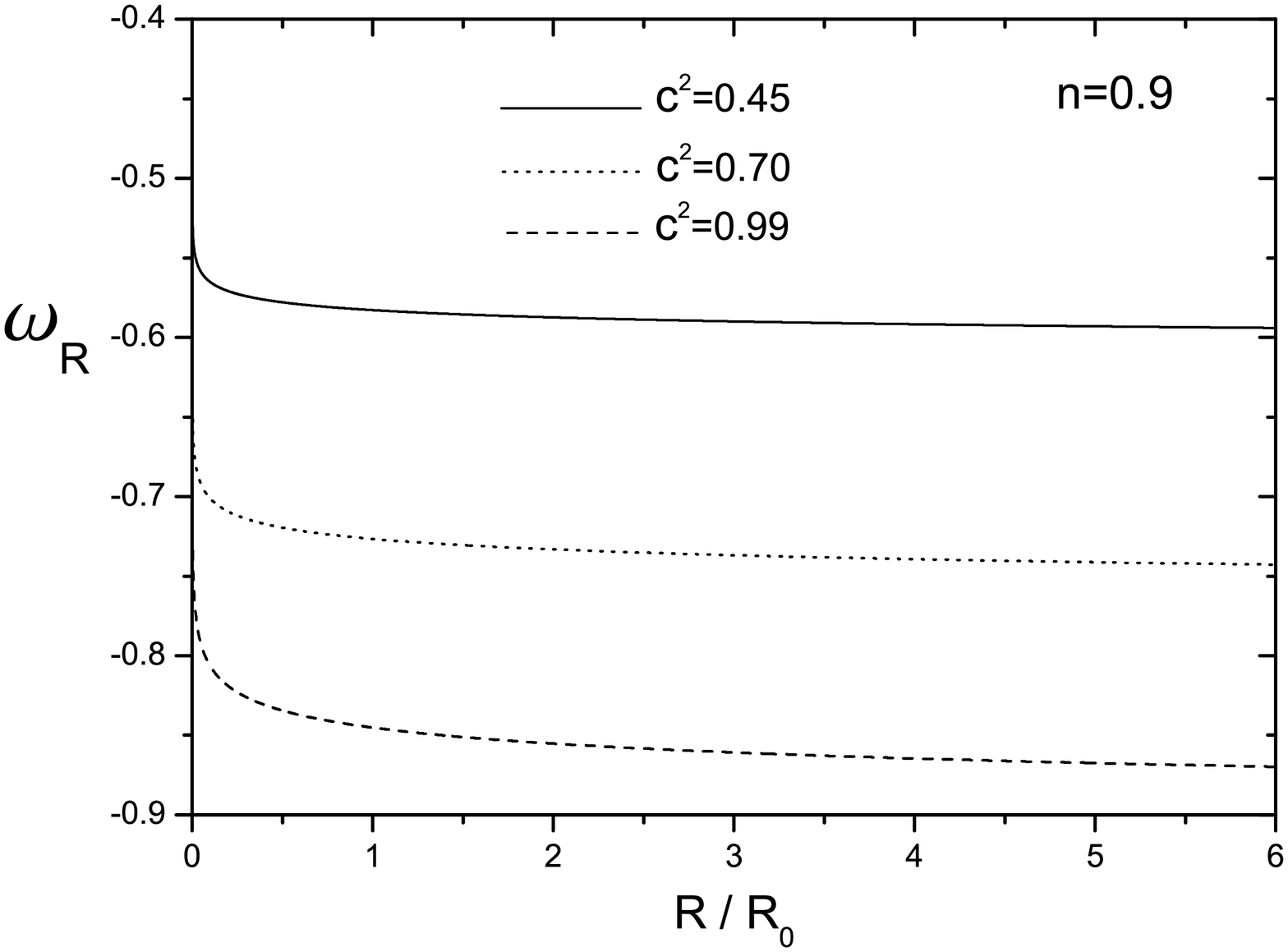}
\caption{Evolution of the effective EoS parameter $w_R$ versus the
dimensionless scalar $R/R_0$, for two different values of the
parameter $n$, in the model $f(R)=\alpha\,R^n$ and
$Q=3c^2H\rho_m$. In both panels, we used three different values of
the interacting-parameter $c^2$. In the upper panel, we have taken
$\alpha=3000$, $n=0.1$ and  the in lower panel we have used
$\alpha=0.09$ and $n=0.9$. Also, in both panels we have used
$H_0=72.5$ Km S$^{-1}$ Mpc$^{-1}$   and $\kappa=1$\label{fig1}}
\end{figure}

In Fig.(\ref{fig1}) we show the evolution from the early times
($R/R_0\rightarrow +\infty$) to the current  epoch ($R/R_0=1$) for
the effective EoS parameter $w_R$ versus the dimensionless scalar
$R/R_0$, for two different values of the parameter $n$ in the
model $f(R)=\alpha\,R^n$. Here, $R_0$ is the Ricci scalar at the
present epoch. In order to write down values that relate the
effective EoS $w_R$ and $R/R_0$, we consider Eq.(\ref{wt2})
together with the interaction term given by Eq.(\ref{QQ}). Here,
we note that we have not a transition from the $w_R>-1$
(quintessence) to $w_R<-1$ (phantom). In this form, the
interaction $Q\propto H\rho_m$ and $f(R)\propto R^n$ gravity
cannot cross the phantom divide line, as could be seen from
Fig.(\ref{fig1}). In both panels, we have used three different
values of the interacting-parameter $c^2$,  where
$\frac{(3-2n)}{3}<c^2<1$.   In the upper panel, we have taken
$\alpha=3000$, $n=0.1$ and in the lower panel we have used
$\alpha=0.09$ and $n=0.9$. Also, in both panels we have used
$H_0=72.5$ Km S$^{-1}$ Mpc$^{-1}$ \citep{rss2009} and $\kappa=1$. From
the upper panel, we note that at the present epoch ($R/R_0=1$),
for the values $c^2=0.94$, $c^2=0.97$ and $c^2=0.99$, we find that
$w_{R}=-0.87$, $w_{R}=-0.93$ and $w_{R}=-0.98$, respectively. Also
we observe that at early times, i.e., $R/R_0\rightarrow\infty$ we
obtain that $w_{R}\rightarrow -0.99$ for all values of $c^2$.
Additionally, we note that the effective EoS parameter $w_R$
depends on $\alpha$ parameter. In particular, for values of
$\alpha<3000$ the effective EoS $w_R\sim -0.99 $ and for values
of $\alpha>3000$, we get that the effective $w_R>0$.

For the case $n=0.9$ (lower panel), we have found that at the
present epoch for the values $c^2=0.45$ $c^2=0.70$ and $c^2=0.99$,
we get that $w_{R}=-0.58$, $w_{R}=-0.73$ and $w_{R}=-0.85$. At
early times, we obtain that $w_{R}\rightarrow-0.65$,
$w_{R}\rightarrow-0.82$ and $w_{R}\rightarrow-0.99$, respectively.
Also, we note that for values of $\alpha<0.09$ the effective EoS
$w_R\sim -0.99$ for the value $c^2=0.99$, $w_R\sim -0.80$ for
$c^2=0.70$ and  for the value of $c^2=0.45$ corresponds to
$w_R\sim -0.65$. For values of $\alpha>0.09$ the effective EoS
$w_R>0$.

\begin{figure}[th]
\includegraphics[width=3.3in,angle=0,clip=true]{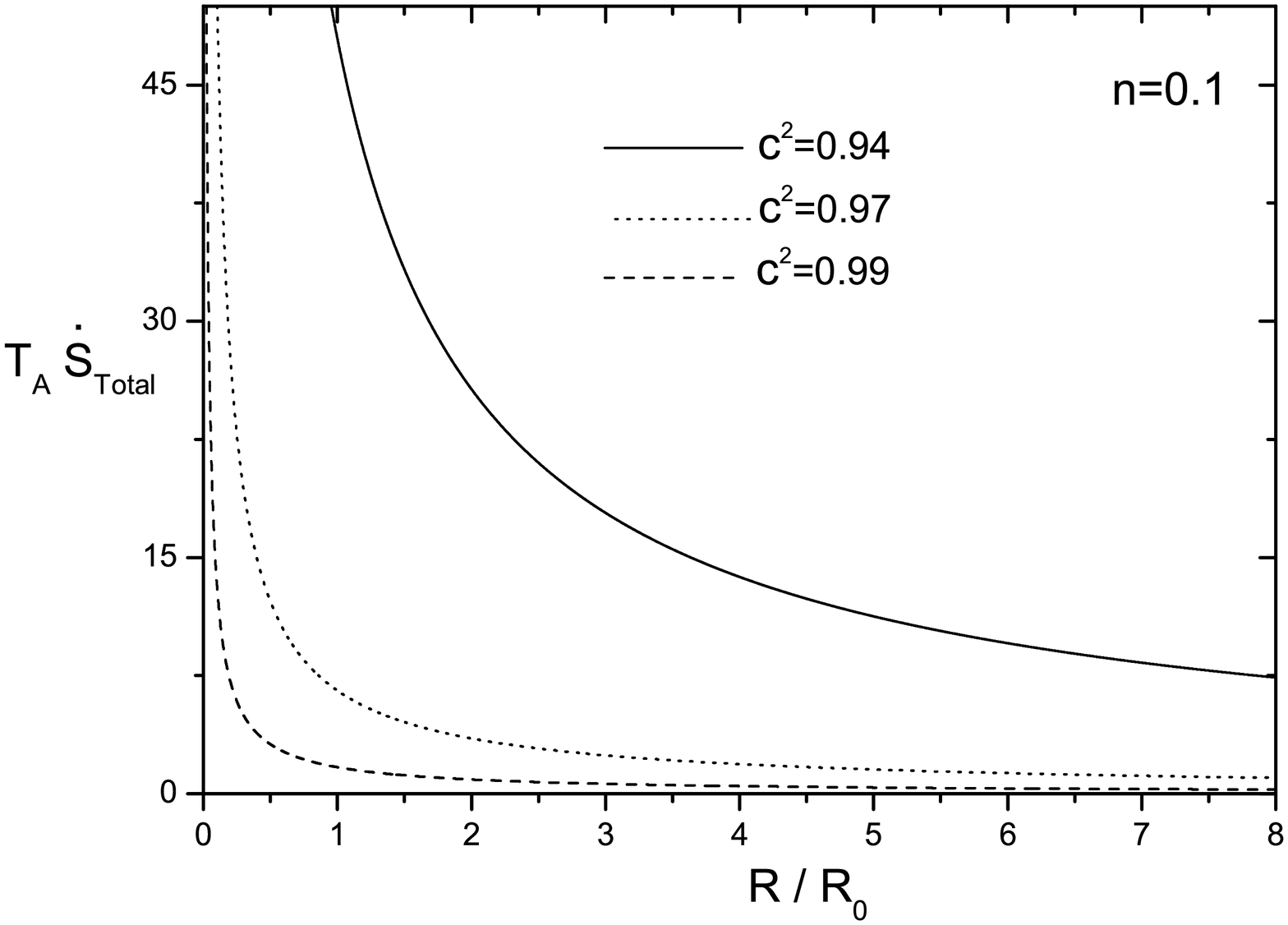}
\includegraphics[width=3.3in,angle=0,clip=true]{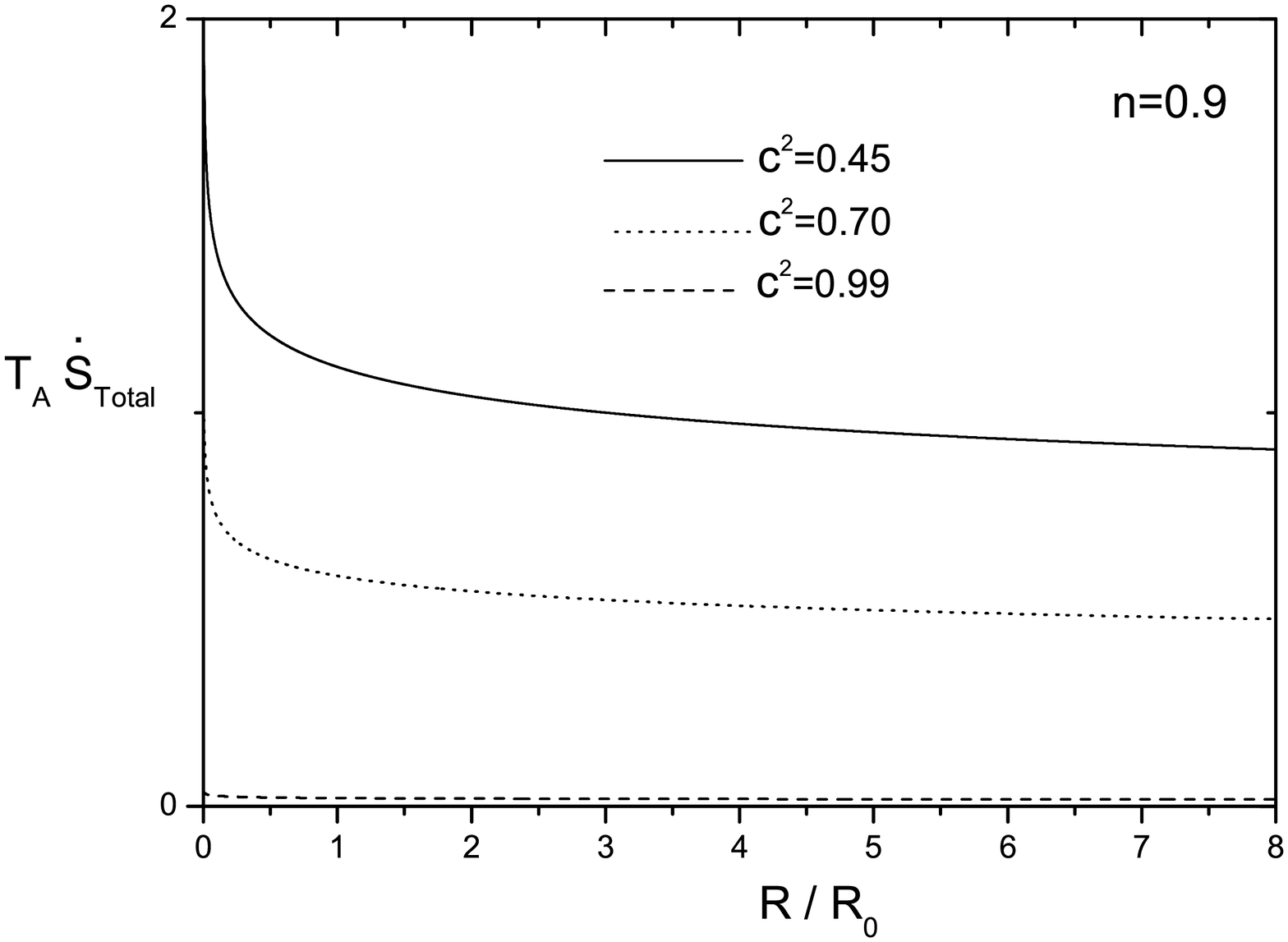}
\caption{Evolution of the GSL ($T_A\dot{S}_{Total}$) versus the
dimensionless scalar $R/R_0$, for two different values of the
parameter $n$. As before, in both panels, we used three different
values of the interaction parameter $c^2$. In the upper panel, we
have taken $\alpha=3000$, $n=0.1$ and in the lower panel we have
used $\alpha=0.09$ and $n=0.9$. Also, in both panels we have used
$H_0=72.5$ Km S$^{-1}$ Mpc$^{-1}$   and $\kappa=1$.\label{fig2} }
\end{figure}

In Fig.(\ref{fig2}) we represent  the evolution  of the GSL versus
the dimensionless scalar $R/R_0$, for two different values of the
parameter $n$.  In order to write down values that relate
$T_A\dot{S}_{Total}$ versus $R/R_0$, we considered Eq.(\ref{ST}).
As before, in the upper panel we have used $\alpha=3000$, $n=0.1$
and in the lower panel we have taken $\alpha=0.09$ and $n=0.9$.
From Fig.(\ref{fig2}) we observe that the GSL is satisfied from
the early times i.e., $R/R_0\rightarrow\infty$ to the current
epoch in which $R/R_0=1$. Also, in both panels we note that the
GSL graphs for the value $c^2=0.99$ corresponds to
$T_A\dot{S}_{Total}\sim 0$. Here, we observe that at early times,
i.e., $R/R_0\rightarrow\infty$ we obtain that
$T_A\dot{S}_{Total}\longrightarrow 0 $ (adiabatic system). In
particular, for $n=0.1$ at the present time i.e., $R/R_0=1$ we get
that $T_A\dot{S}_{Total}\simeq 47.9 $ for the value $c^2=0.94$,
$T_A\dot{S}_{Total}\simeq 6.6$ that corresponds to $c^2=0.97$ and
$T_A\dot{S}_{Total}\simeq 1.7$ that corresponds to $c^2=0.99$.
For the specify  case $n=0.9$ at the present time ($R/R_0=1$),  we
find  that $T_A\dot{S}_{Total}\simeq 1.12 $ for the value
$c^2=0.45$, $T_A\dot{S}_{Total}\simeq 0.59$ that corresponds to
$c^2=0.70$ and $T_A\dot{S}_{Total}\simeq 0.02$ that corresponds to
$c^2=0.99$. Also, we note that the GSL is increased  in the future
i.e., $0<R/R_0<1$, in which
 $T_A\dot{S}_{Total}>0$.

\section{Conformal transformation: Scalar tensor gravity theory}\label{co}

 Since that there are two approaches  to study the dark energy
$f(R)$-gravity model, is interesting to analyze the GSL of our
model as a scalar tensor gravity theory, by means of  a conformal
transformation,  specifically from the original frame (called
Jordan frame) to the Einstein frame. Introducing a conformal
transformation, the metric tensor $g_{\mu\nu}$ is transformed into
$\widetilde{g}_{\mu\nu}= \Omega(x)^2g_{\mu\nu}$, where
$\Omega(x)^2$ is the conformal factor, $g_{\mu\nu}$ and
$\widetilde{g}_{\mu\nu}$ represent the original and transformed
metric, respectively. With this conformal transformation and
together with the introduction of new field $\sigma$, defined as
$$
\Omega(x)^2=e^{\sqrt{\frac{2}{3}}\,\sigma}=f_R,
$$
the action given by Eq.(\ref{action}) becomes a Einstein Hilbert
type action \cite{Am}, given by
$$
I_E=\int\,d^4x\,\sqrt{-\widetilde{g}}\,\left[\frac{R(\widetilde{g})}{2}-\frac{1}{2}(\widetilde{\nabla}\sigma)^2-V(\sigma)+\widetilde{L}_m\right],
$$
where the scalar field potential, becomes \cite{Am}
$$
V(\sigma)=\mbox{sign(F)}\;\;\frac{(f-Rf_R)}{2f_R^2}.
$$
For the case in which the metric in the new coordinates,
corresponds to FRW metric, then  the relation between the scale
factor $\widetilde{a}\equiv a_E$ in the Einstein frame and the
scale factor in the Jordan frame, is given by\cite{Am}
\begin{equation}
a_E=e^{\frac{\sigma}{\sqrt{6}}}\,a,\label{aae}
\end{equation}
and the time coordinate in the Einstein frame $\widetilde{t}\equiv
t_E$ and the time in the Jordan frame are related by the
differential relationship \cite{Am}
\begin{equation}
e^{\frac{-\sigma}{\sqrt{6}}}\,dt_E=dt,\label{tt}
\end{equation}
also, the transformation for the energy density of the matter in
both frames, becomes
$$
\widetilde{\rho}_m\equiv\rho_{m_E}=\rho_m\,
e^{-2\sqrt{\frac{2}{3}}\,\sigma}.
$$
In the Einstein frame, the energy density $\rho_E$ and the scalar
field $\sigma$ satisfy the following equations:
\begin{equation}
\rho_{m_E}'+3H_E\rho_{m_E}+\sqrt{\frac{1}{6}}\,\sigma'\,\rho_{m_E}=\widetilde{Q}=
e^{-\frac{5}{\sqrt{6}}\,\sigma}\,Q,\label{crm}
\end{equation}
\begin{equation}
\left[\sigma''+3H_E\sigma'+V,_\sigma-\sqrt{\frac{1}{6}}\,\,\rho_{m_E}\right]\sigma'=-\widetilde{Q}.\label{cp}
\end{equation}
Here, we noted that in the Einstein frame appears an effective
interaction term
$Q_{eff}=\widetilde{Q}-\frac{1}{\sqrt{6}}\rho_{m_E}\,\sigma'$, and
in the limit $\widetilde{Q}\rightarrow 0$ (or analogously
$Q\rightarrow 0$), then $Q_{eff}\rightarrow
-\frac{1}{\sqrt{6}}\rho_{m_E}\,\sigma'$.

The Friedmann equation, in this frame, results
\begin{equation}
3H_E^2=\frac{\sigma'^2}{2}+V(\sigma)+\,\rho_{m_E},\label{cH}
\end{equation}
where the primes denote differentiation respect to the time $t_E$
and $H_E=a_E'/a_E$ defines the Hubble parameter in the Einstein
frame.

From Eqs.(\ref{Sm}), (\ref{SA}),(\ref{aae}) and (\ref{tt}), the
GSL of thermodynamics due different contributions of the apparent
horizon entropy and the matter entropy in the Einstein frame, can
be written as
$$
T_A\,S'_{Total}=4\pi\Bigl[
R_A^2\,\rho_{m_E}\,e^{\sqrt{3/2}\sigma}\left[e^{\sqrt{1/6}\sigma}R_A'+\left(\frac{\widetilde{Q}}{3\rho_{m_E}(H_E-\sigma'/\sqrt{6})
}-1\right)\right]+
$$
\begin{equation}
e^{\sqrt{2/3}\sigma}\left(1-\frac{e^{\sqrt{1/6}\sigma}\,R_A'}{2}\right)\,\left(2R_A'+\sqrt{\frac{2}{3}}\,\sigma'\,R_A\right)
\Bigr],\label{ES}
\end{equation}
where now the radius of the apparent horizon in the Einstein
frame, is
$$
R_A=\frac{e^{\sigma/\sqrt{6}}}{(H_E-\sigma'/\sqrt{6})}.
$$
In the following, we will study the GSL of thermodynamics for our
specific model. Considering the case in which $f(R)$ is given by
Eq.(\ref{fT2}) i.e., $f=\alpha\,R^{n}$  and $Q$  by Eq.(\ref{QQ}),
we get that the scalar field potential is
$$
V(\sigma)\sim\, \exp\left[-\lambda\,\sigma\right],
$$
where the constant
$\lambda=\sqrt{\frac{2}{3}}\,\,\left[\frac{2-n}{1-n}\right]$.

From the new equations of motion, the solution in the Einstein
frame for the energy density of the matter, becomes $
\rho_{m_E}\sim\, t_E^{-2} $, and  the solution for the scalar
field $\sigma$, is given by
\begin{equation}
 \sigma=\frac{2}{\lambda}\,\ln(t_E). \label{ae}
\end{equation}
The scale factor in the Einstein frame, by using
 Eqs. (\ref{sol1}),(\ref{aae}) and (\ref{ae}), becomes
\begin{equation}
a_E\propto\,t_E\,^\gamma\,,\,\;\,\;\;\mbox{where}\,\,\,\,\;\;\gamma=\frac{2}{\sqrt{6}\,\lambda}\,+\left(1-\frac{2}{\sqrt{6}\,\lambda}\right)\,\frac{2n}{3(1-c^2)},
\end{equation}
where the exponent in the scalar factor is $\gamma>1$ for
guarantee an accelerated phase of the universe. Using the fact
that $0<n<1$, together with the condition $\gamma>1$, we find that
the range for the parameter $c^2$, becomes
$\frac{(3-2n)}{3}<c^2<1$, that is similar to obtained in the
Jordan frame.

The Hubble parameter in the Einstein frame is given by
$$
H_E=\frac{1}{a_E}\frac{da_e}{dt_E}=\frac{\gamma}{t_E}=\sqrt{\frac{\gamma}{6(2\gamma-1)}\,R_E},
$$
where $R_E$ represents the scalar curvature  in the Einstein
frame.

\begin{figure}[th]
\includegraphics[width=3.3in,angle=0,clip=true]{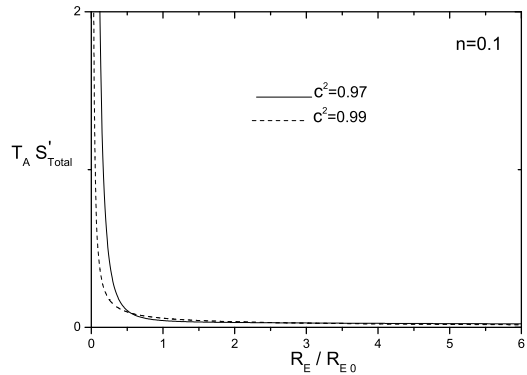}
\caption{Evolution of the GSL ($T_A\,S'_{Total}$) versus the
dimensionless scalar $R_E/R_{E\,0}$ in the Einstein frame, for two
different values of the parameter $c^2$ in the case
$n=0.1$.\label{fig3} }
\end{figure}

In Fig.(\ref{fig3}) we represent  the evolution  of the GSL versus
the dimensionless scalar ratio $R_E/R_{E0}$ in the Einstein frame,
for the case $n=0.1$ and  for two different values of the
parameter $c^2$. In order to write down values that relate
$T_A\,S'_{Total}$ versus $R_E/R_{E0}$, we used Eq.(\ref{ES}),
together with our specific case, i.e., $f(R)=\alpha\,R^{n}$ and
$Q=3c^2\,H\,\rho_m$.  From Fig.(\ref{fig3}) we observe that the
GSL is satisfied from the early times i.e.,
$R_E/R_{E0}\rightarrow\infty$ to the current epoch in which
$R_E/R_{E0}=1$.  Also, we find that at early times, i.e.,
$R_E/R_0\rightarrow\infty$, the GSL
$T_A\,S'_{Total}\longrightarrow 0 $,  for both values of $c^2$. In
particular, for $n=0.1$ at the present time i.e., $R_E/R_{E0}=1$,
we get that $T_A\,S'_{Total}\simeq 4.3\times 10^{-2} $, for the
value $c^2=0.97$, and  $T_A\,S'_{Total}\simeq  5.8 \times 10^{-2}
$ that corresponds to $c^2=0.99$. Also, we noted that for values
of the parameter $c^2<0.965$, the GSL of thermodynamics is
negative, $T_A\,S'_{Total}<0$, and then the GSL is violated for
the dimensionless scalar ratio $R_E/R_{E0}$. For the other specify
case $n=0.9$, we find  that $T_A\,S'_{Total}<0$, in which the GSL
is violated for the values $c^2<0.977$ (figure not shown). In this
form, we noted that the validity of the  GSL of thermodynamics in
both frames is non equivalent.

\section{Conclusions \label{conclu}}

In this paper we have investigated the GSL in the context of
interacting $f(R)$ gravity. We studied the GSL from the boundary
of the universe to be enclosed by the dynamical apparent horizon
in a flat FRW universe occupied with pressureless DM, together
with the Hawking temperature on the apparent horizon. We have
found that the interacting term $Q$ modified; the curvature
contributions component given by an effective EoS parameter, the
deceleration parameter and the  evolution of the total entropy or
rather the GSL. In particular, we have obtained  that the
modification in the evolution of the total entropy, results in an
increases on the GSL of thermodynamics by a factor
$4\pi\,R_A^3\,Q/3>0$.

Our specific model  is described by a  model $f(R)\propto R^n$ and
we have considered for simplicity the case in which the
interaction term $Q$ is related to the total energy density of
matter. For this specify  model, we have  found analytic solutions
and obtained explicit expressions for the effective EoS parameter,
the deceleration parameter and the evolution of the total entropy.
For this model, we observed that we do not have a transition from
the value $w_R>-1$ (quintessence) to $w_R<-1$ (phantom) and  the
interacting $f(R)=\alpha R^n$ gravity cannot cross the phantom
divide line, as could be seen from Fig.(\ref{fig1}). Also, we have
observed that the GSL is satisfied from the early times i.e.,
$R/R_0\rightarrow\infty$ to the future in which GSL always
$T_A\dot{S}_{Total}>0$.

Also, we have shown that the GSL of thermodynamics for the
interacting  $f(R)$ gravity is less restricted than analogous
$Q=0$ due to the introduction of a new parameter, present in the
interaction term $Q$. In our specific model the
incorporation of this parameter gives us a freedom that allows us
to modify the standard $f(R)$ gravity by simply modifying the
corresponding value of the parameter $c^2$.

 We have studied the GSL of thermodynamics for the interacting
$f(R)$  in the Einstein frame through a conformal transformation.
In particular, for our specific model in which  $f(R)\sim R^n$ and
$Q\sim c^2\,H\,\rho_m$, we have observed that the GSL
thermodynamic is violated in the Einstein frame for  some values
of the parameters $n$ and $c^2$. In particular, for $n=0.1$ we
have found that for values of $c^2<0.965$, the GSL is negative,
$T_A\,S'_{Total}<0$, and then the GSL is violated for the
dimensionless ratio $R_E/R_{E0}$. Similarly, for $n=0.9$ we have
obtained that $T_A\,S'_{Total}<0$, in which the GSL of
thermodynamic is violated for the values of $c^2<0.977$. In this
form, we have found that the validity of the  GSL of
thermodynamics in both frames is non equivalent.

Finally, we have not addressed other interacting-$f(R)$ models (see
e.g., Refs. \citep{jai2013, mira2009, kawa2008, call2004, nji2003, card2003, salva2002}). Here, a more accurate numerical
calculation would be necessary for different  $f(R)$ gravity models and
$Q$ interaction terms. We hope to return to this point in near
future.

\begin{acknowledgments}
R.H. was supported by COMISION NACIONAL DE CIENCIAS Y TECNOLOGIA
through FONDECYT grant  N$^0$ 1130628  and by DI-PUCV grant
123.724. N.V. was supported by Proyecto Beca-Doctoral CONICYT
N$^0$ 21100261.
\end{acknowledgments}

\end{document}